# Deep Learning Decoding of Mental State in Non-invasive Brain Computer Interface

Dongdong Zhang
Institute of Brain Science
DeepBlue Academy of Sciences
DeepBlue Technology
(Shanghai)Co., Ltd.
zhangdongdong@deepblueai.com

Dong Cao
Institute of Cognitive Intelligence
DeepBlue Academy of Sciences
DeepBlue Technology
(Shanghai)Co., Ltd.
caodong@deepblueai.com

Haibo Chen
DeepBlue Academy of Sciences
DeepBlue Technology
(Shanghai)Co., Ltd
chenhaibo@deepblueai.com

## ABSTRACT

Brain computer interface (BCI) has been popular as a key approach to monitor our brains recent year. Mental states monitoring is one of the most important BCI applications and becomes increasingly accessible. However, the mental state prediction accuracy and generality through encephalogram (EEG) are not good enough for everyday use. Here in this paper we present a deep learning-based EEG decoding method to read mental states. We propose a novel 1D convolutional neural network with different filter lengths to capture different frequency bands information. To improve the prediction accuracy, we also used a resnet-like structure to train a relatively deep convolutional neural network to promote feature extraction. Compared with traditional ways of predictions such as KNN and SVM, we achieved a significantly better result with an accuracy of 96.40%. Also, in contrast with some already published open source deep neural network structures, our methods achieved the state of art prediction accuracy on a mental state recognition dataset. Our results demonstrate using only 1D convolution could extract the features of EEG and the possibility of mental state prediction using portable EEG devices.

• Human-centered computing → Human computer interaction (HCI) • Applied computing → Life and medical sciences

## 1 INTRODUCTION

Brain computer interface (BCI) is an electronic device that establish a communication channel between our biological brain and external equipment [1]. BCI is becoming a technology that will change our lives dramatically and plays a central role in reading our thoughts and transfers the information to the outside world [2]. One of the most important applications of non-invasive BCI is brain state monitoring through electroencephalogram (EEG) or near infrared spectrum (NIRS) [3]. EEG has a long history and has been adopted for brain signal decoding or brain diseases diagnosis for more than 100 years [4]. As the rise the BCI technology, EEG quickly becomes one of the main brain signal acquisition techniques because of its ease of use.

Previous studies have employed EEG to read people's attention [5]. Liu et al reported that they trained a support vector machine (SVM) classifier from EEG data to distinguish students' attentiveness and achieved an accuracy of 76.82% [6]. Wang et al showed that they could read the engagement of operators which indicated their attentions in dual (multi)-tasking conditions through EEG recording [5].

Some studies also showed the possibility of using EEG to detect fatigue in humans. The driving fatigue is disastrous and can cause serious traffic accidents. EEG activity changes is the most direct manifestation of fatigue [7] and combining EEG and EOG as a biomarker achieves high prediction accuracy of fatigue [8]. The other study demonstrated mental fatigue was associated with increased power in frontal theta (θ) and parietal alpha (α) EEG rhythms [9].

There are also some studies showed probability of monitoring mental state using EEG. Drowsiness detection proved to be promising and 92% prediction accuracy was achieved in one study [10]. Richer et al built a entropy-based measure to recognize Focus and Relax states in real-time with over 80% accuracies [11]. Acl et al demonstrated the power of EEG-based passive BCI and its ability to distinguish mental attention states [12].

Deep learning has been around for almost 10 years and revolutionized nearly all the fields in artificial intelligence [13]. Several studies have started to use deep learning techniques to decode brain activities in BCI applications [14]. Roy et al showed the potential of using deep recurrent neural network to diagnose brain-related disorders and proposed a RNN architecture called ChronoNet [15]. Sakhavi et al proposed a classification framework for motor imaginary (MI) data by introducing a new temporal representation of the data and also utilizing a convolutional neural network (CNN) architecture for classification. The performance of this framework outperformed the best method by 7% [16]. These studies demonstrate the great potential of deep learning in the future applications of BCI.

In this study, we proposed a structure using only 1D convolution neural network to predict mental state collected from a wearable EEG device. We designed filters to extract different bands features and achieved state of art performance.

## 2  METHODS

All the experiments in this paper was done with the data from this study [12]. The dataset primarily comprised a total of 25 hours EEG recordings collected from 5 participants engaged in a low-intensity control task. The task consisted of controlling a computer-simulated train. Each experiment lasted for 35 to 55 min over a primarily featureless route.

### 2.1 Experiment procedure

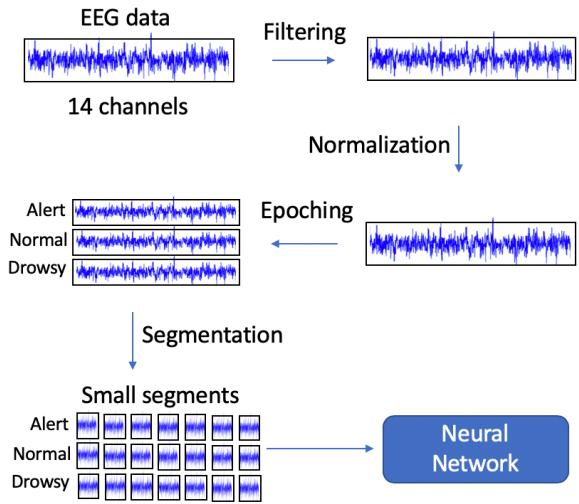

**Figure 1: Framework of data processing workflow**

Detailed experiment procedure was described here [12]. The dataset contains 31 recording sessions from 5 subjects. Different mental states were induced in a driving simulation environment. There are 3 mental states recorded in this dataset including the focused, the normal (unfocused or detached) state and the drowsy state. Participants were asked to control a simulated passenger train for a duration of 35 to 55 minutes during each recording session. During the first 10 minutes of each session, the participants were asked to remain focused to control the train. During the second 10 minutes, the participants were in a relax state without controlling the train anymore. Finally, during the last 10 minutes, the participants were allowed to close their eyes or even doze off as desired.

### 2.2 Data preprocess

EEG data were collected using Emotiv Epoc+ headset. This headset has 14 recording channels at a sampling rate of 128 Hz and a bandwidth between 0.2-43 Hz. First, the raw EEG data was filtered with a bandpass filter with cutoff frequencies of 1 and 40 Hz. The epoch of each mental state was approximately 10 minutes. There were 31 recording sessions which were randomly divided into 5 groups and we adopted a 5-fold cross validation recipe. We split the data to get the epochs for the three distinct mental states. Then, each mental state was segmented into 15 s clips with 14 s overlap, which was similar to the parameter settings described in Acı et al feature extraction procedure. Additionally, we normalized the data as follows:

$$x_{i,j} = \frac{x_{i,j} - \overline{x}}{s}$$

$x_{i,j}$ is data point $x_i$ in channel j, $\overline{x}$ is the mean of all the channel data, $s$ is the standard deviation of the data. We performed the normalization using total average and deviation to preserve the correlations between different channels.

We adopted two kinds of cross validation methods described as follows. First, to compare our approach with traditional methods used in Acı et al, we kept the same cross validation procedure in which we segmented the data into small overlapped clips and then randomly split into training and validation sets. In this case, there is a high probability that the validation folds could overlap with the training folds which may cause high biases in our result (table 1). Therefore, we modified this procedure in the deep net comparisons. To ensure the exclusiveness of the validation set, we first randomly selected 6~7 independent recording sessions as our validation set before segmentation. The rest 34~35 sessions were marked as training set (table 2).

### 2.3 Structure of deep convolutional neural network

Deep neural network (DNN) has been successfully and widely used in speech recognition, image recognition and understanding, natural language processing. Until recently, the area of BCI starts embracing this technique and a lot of progress has been made. EEG signals are time series and tricks used in audio processing can be applied to EEG analysis immediately.

In this work, we constructed a deep neural network comprising of the input layer, 6 convolution blocks, an output layer as shown in Figure 2. In each convolution block, two consecutive 1-D convolution with batch normalization and nonlinear relu activation was followed by one shortcut connection. A down sampling layer was added after the shortcut connection to promote generalization. By employing shortcut connections, we could train a comparably deep neural network with better performance. The filter length is constant in one block. At the end of convolution block, a dropout layer with 0.5 keeping probability was added to reduce overfitting. The final softmax layer was used for the final mental state classification. We fit the model using the Adam optimizer with default parameters. Cross-entropy has been used for the loss function.

All models were trained on a 5 Nvidia 1080 Ti GPUs workstation with CUDA 9.0 using the Keras API with Tensorflow backend.

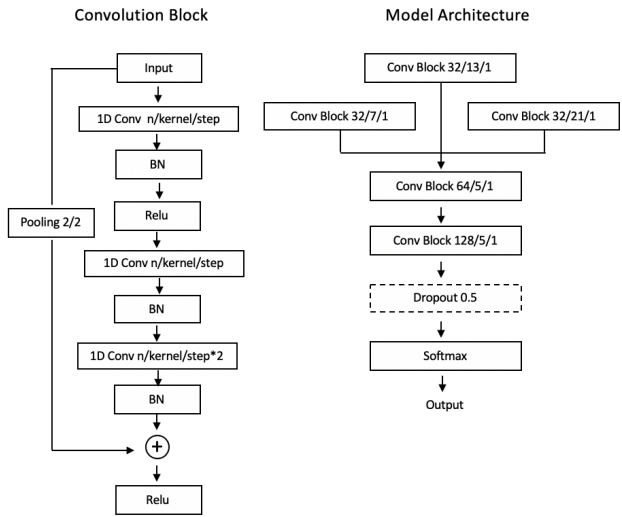

**Figure 2: Visualization of DNN structure**

## 3 RESULTS

Experiments of mental state prediction was performed on the preprocessed dataset. Previous EEG studies tended to train one classifier for each subject because of the poor generalization of the algorithms, which is also called subject-specific paradigm. A lot of such classifiers achieve good performances but it's impossible to train a detector for each individual in real life. Therefore, we adopted a more general method and trained a universal classifier for all the subjects, which is also called common-subject paradigm. We directly trained a 3-classes deep neural network and the 5-fold cross validation technique was applied to evaluate the performance.

Table 1 showed the prediction accuracies results based on our trained neural network. It shows that our model can predict much better than the traditional algorithms such as support vector machine, adaptive neuro-fuzzy system, or k-nearest neighbors. Our proposed method achieved an accuracy of 96.40%.

We also tried to compare our method with some published open source deep learning architectures as shown in table 2. EEGNet is a compact convolutional neural network using depthwise and separable convolutions to construct a model to extract EEG features. FBCSPNet employed the idea of widely used filter bank common spatial patterns (FBCSP) algorithm to build a deep neural net. Results showed that our model is better than those methods.

**Table 1: Comparisons of the prediction accuracy with traditional methods**

| Methods | Prediction Accuracy |
| --- | --- |
| kNN (Acl et al [12]) | 77.76% |
| ANFIS (Acl et al [12]) | 81.55% |
| SVM (Acl et al [12]) | 91.72% |
| Our proposed | 96.40% |

**Table 2: Comparisons of the prediction accuracy with other deep learning models**

| Methods | Prediction Accuracy |
| --- | --- |
| EEGNet ([17]) | 51.01% |
| FBCSPShallowNet ([18]) | 49.17% |
| DeepConvNet ([18]) | 52.91% |
| Our proposed | 53.22% |

## 4 DISCUSSION

In this work, we examined the problem of mental state prediction from EEG data using deep learning algorithms. By comparing different EEG bands power to recognize mental state has been popular in many studies [19]. A lot of preprocessing tricks and machine learning algorithms has been proposed to overcome the susceptibility of EEG BCI to motion artifacts or external electromagnetic interferences [20]. Different from those studies, we adopted convolution neural network which is an automatically powerful feature extractor to distill useful features. The automatic feature extraction procedure is superior to traditional method by hand picking as it keeps all the useful information in the original data [17].

Some studies suggested using recurrent neural network (RNN) to extract temporal information of EEG data [15]. This is true in some circumstances. However, due to the slow computation of RNN, it could hinder its usage in small chips with less computing power. Studies have demonstrated that 1-D convolution could also extract temporal information [16]. We used 1-D convolution layers only in our deep neural network architecture without any RNN to facilitate training and inference. Compared with some published deep neural nets [17], [18], our method get a higher prediction accuracy proving only 1D convolution without CNN is good enough to extract EEG features.

To compare our approach with traditional methods used in Acı et al, we kept the same cross validation procedure in which we segmented the data into small overlapped clips and then randomly split into training and validation sets. This validation approach allowed leakage of information between the training and validation sets. Such cross-validation cannot be used as a reliable performance metric. Therefore, we abandoned such approach and randomly split training and validation sets before segmentation to avoid information overlap. The change of cross-validation strategy significantly decreased the model performance but yielded a more realistic and objective evaluation.

The present work showed that deep learning could significantly improve the prediction accuracy and generality. Although the method proposed achieved a better result, the prediction accuracy 53.22% is still relatively low. As the data was collected from other studies and the amount of the data is small, we may try to collect more data to verify the power of this architecture in the future. The normal state is hard to predict because it is an intermediate condition that could be easily misclassified (data not shown). Besides the mental state data, we may try other dataset such as driving fatigue or motor imaginary to explore potential usages. All the recording channel were included in this study. A good channel selection scheme is needed to reduce the information redundancy and to achieve equal prediction accuracy with less computing power.

## ACKNOWLEDGMENTS

We are very grateful to DeepBlue Technology (Shanghai) Co., Ltd. and DeepBlue Academy of Sciences for their support.